# Edge-Of-Chaos Learning Achieved by Ion-Electron Coupled Dynamics in an Ion-Gating Reservoir

Daiki Nishioka[1,2], Takashi Tsuchiya[1]*, Wataru Namiki[1], Makoto Takayanagi[1,2], Masataka Imura[3], Yasuo Koide[4], Tohru Higuchi[2], and Kazuya Terabe[1]

[1]International Center for Materials Nanoarchitectonics (WPI-MANA), National Institute for Materials Science (NIMS), 1-1 Namiki, Tsukuba, Ibaraki, 305-0044, Japan.

[2]Department of Applied Physics, Faculty of Science, Tokyo University of Science, Katsushika, Tokyo 125-8585, Japan

[3]Research Center for Functional Materials, NIMS, 1-1 Namiki, Tsukuba, Ibaraki, 305-0044, Japan.

[4]Research Network and Facility Services Division, NIMS, 1-2-1 Sengen, Tsukuba, Ibaraki, 305-0047, Japan.

*Email: TSUCHIYA.Takashi@nims.go.jp

## Abstract

Physical reservoir computing has recently been attracting attention for its ability to significantly reduce the computational resources required to process time-series data. However, the physical reservoirs that have been reported to date have had insufficient expression power, and most of them have a large volume, which makes their practical application difficult. Herein we describe the development of a $Li^+$-electrolyte based ion-gating reservoir (IGR), with ion-electron coupled dynamics, for use in high-performance physical reservoir computing. A variety of synaptic responses were obtained in response to past experience, which responses were stored as transient charge density patterns in an electric double layer, at the $Li^+$-electrolyte/diamond interface. Performance, which was tested using a nonlinear autoregressive moving-average (NARMA) task, was found to be excellent, with a NMSE of 0.023 for NARMA2, which is the highest for any physical reservoir reported to date. The maximum Lyapunov exponent of the IGR was 0.0083: the edge of chaos state enabling the best computational capacity. The IGR described herein opens the way for high-performance and integrated neural network devices.

## Introduction

Artificial neural network (ANN)-based information processing is becoming more and more important as a way to deal with the vast amount of information currently in existence [1, 2]. ANN computing [e.g., deep learning with a multi-layer neural network (NN)] can provide excellent learning, classification, and inference characteristics that are close to, and in some cases beyond, those found in natural intelligence (i.e., the human brain), whereas the enormous amounts of power required by ANN (as in a typical multi-layer NN) are far higher than that required by human beings [2]. The low energy efficiency of ANN computing is a serious drawback in the realization of ubiquitous and versatile AIs, but is inherent in the structure of the ANN, which require the weights of millions of virtual synaptic nodes to be stored and updated (i.e., large network size) and requiring the consumption of tremendous amounts of energy. Reservoir computing (RC) has recently been attracting attention because of its ability to significantly reduce the computational resources required to process time-series data, which it is able to do because of its utilization of the nonlinear responses of a 'reservoir' to input signals. While simulated recurrent NNs have been used as reservoirs to perform fully-simulated RC [3-5], materials or devices with nonlinearity, high-dimensionality and short-term memory have been explored as possible 'physical reservoirs' that can process information without heavy computational burdens for complicated simulations of the dynamical states of a reservoir [6]. To date, the nonlinear dynamics of various materials and devices (e.g., soft bodies, optical devices, spin torque oscillators, and memristors) have been reported as providing nonlinear dynamics that are sufficient to perform physical reservoir-based RC with various time-series tasks, including image recognition, spoken digit classification, and combinatorial optimization [6-24]. However, to date, the performance of physical reservoir-based RC has been far from satisfactory due to the low expression power of physical reservoirs in comparison to the RC performance of simulated reservoirs. Furthermore, most of the high performance physical reservoirs have large volumes, over several $cm^{-3}$, which are not realistic choices for practical application to integrated AI devices [7-9]. Therefore, achieving compatibility between (i) the high expression power of a physical reservoir and (ii) small reservoir volume is a great challenge in nanotechnology research leading towards the physical implementation of RC at practical levels.

Herein we report the achievement of high performance physical RC using an ion-gating reservoir (IGR), in which ion-electron coupled dynamics at a lithium ion electrolyte/diamond interface generate an 'edge-of-chaos' state, which is empirically known to exhibit high computational performance [25]. Various synaptic responses, with asymmetric relaxation and spikes, were obtained with respect to the input history of a single IGR transistor (IGRT), which operates in an electric double layer (EDL) mechanism[26-33], to achieve excellent expression power in a physical reservoir-based RC. Furthermore, a strong dependence of the synaptic response on channel length, which is a feature of ion-electron coupled dynamics, is utilized to realize high dimensionality in a single IGRT. The IGRT exhibited small errors in some RC tasks, including 0.023 of NMSE in a NARMA task, which is a typical bench mark for RC [7,8,13,14,34-37], and achieved 88.8% accuracy in a hand-written digit recognition task. The underlying mechanism of the characteristic synaptic response was investigated on the basis of multiphysics simulation, and it was found that complexed charge density patterns form and change from moment to moment in an extremely thin EDL region (<2 nm) during storage and processing of input signals. We further performed a Lyapunov analysis to investigate a possible origin of the high performance from a nonlinear dynamics viewpoint. The calculated value of the maximum Lyapunov exponent was $8.3 \times 10^{-3}$, close to an 'edge of chaos' state, which is located between order ($\lambda$ <0) and chaos ($\lambda$ >0) and is empirically known to derive high expression power from a reservoir in RC [14,22,24,38-40]. Due to the many advantages observed, including (i) the very thin nature of EDL (e.g., nm order) and the spontaneous formation at interfaces, and (ii) the strong nonlinear response based on ion-electron coupled dynamics, our approach is useful for realizing high performance, highly integrated, and low power consumption AI devices by harnessing the inherent physical and chemical characteristics of materials.

## Results

### Synaptic responses of IGRT and its application to image recognition

In biological neuronal network systems, synaptic responses show characteristic variations in waveform, intensity, and frequency, with respect to environmental inputs in various forms due to the chaotic dynamics of the neural network, as illustrated in Fig. 1a [41,42]. The wide variations in synaptic responses are utilized to achieve high expression power for efficient information processing. In this study, we employ an all-solid-state ion-gating transistor, operating in an electric double layer (EDL) mechanism, in order to mimic the above mentioned wide variation in synaptic response. As shown in Fig. 1a, the transistor consists of a lithium ion conducting solid electrolyte (Li-Si-Zr-O) and a hydrogen-terminated diamond (100) single crystal with homoepitaxial layer, which works in the manner of an EDL transistor [33]. Said transistor is utilized as an ion-gating reservoir (IGR), which is a novel class of physical reservoirs. The ion-gating reservoir can map time-series data in high dimension feature space by using $I_D$ response, with asymmetric relaxation and spikes, the characteristics of which are widely modified by the input history. To investigate a function of IGR as a physical reservoir, we performed a handwritten-digit recognition task [13,16,18,43]. Fig. 1b is an example of the $28 \times 28$ pixel input digit '6' from the MNIST database [44]. Said image was converted into binary time series data and input to the IGRT. The reservoir states were obtained from $I_D$. Fig. 1c shows the 16 different reservoir states, which were well separated from each other so that all 16 different pixel combinations could be expressed by unique reservoir states. These values were used as the reservoir output to train and test the readout network. Similar methods have been used elsewhere [13,16,18,43]. The details of the procedure employed are given in the Methods section. Fig. 1d shows image recognition accuracy vs. the number of trained images. The recognition accuracy improved from 70.4% to 88.8% as the number of trained images increased from 100 to 60 thousand, supporting understanding that the recognition task is suitable for the IGR. While the performance was not as good as that achieved by a typical 3-layer neural network(NN), the size of the network in the present study (1,960) is far smaller than in a 3-layer NN(784,000). Compared to the recognition accuracies of other physical reservoirs (83% to 90.2%) [13,16,18,43], that of IGR is similar or slightly better. However, while the advantage of IGR is minor for such a relatively easy task, IGR showed very good computational performance on more difficult time series data analysis tasks that require superior properties, such as reservoir diversity, which is discussed below.

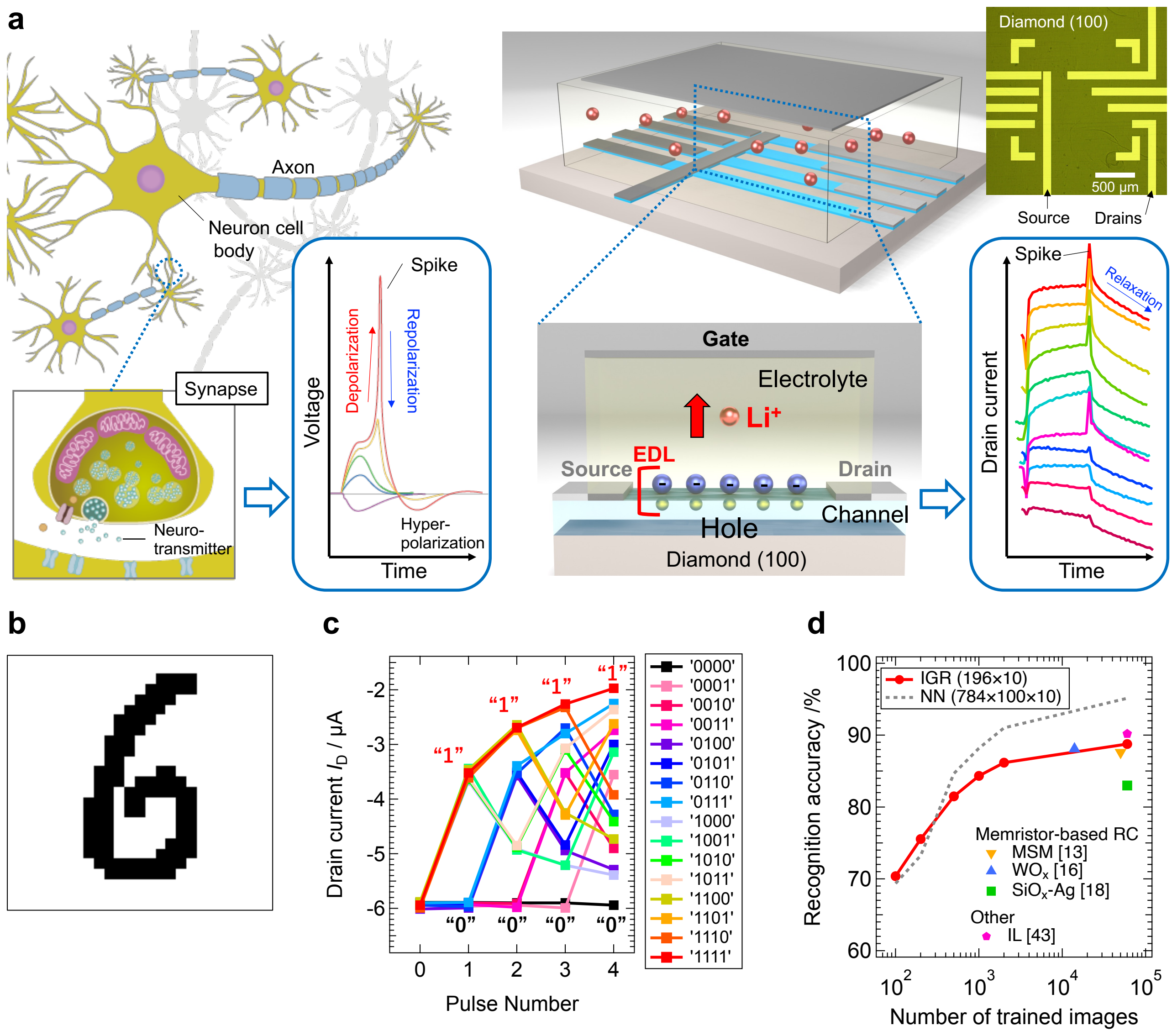

**Fig. 1 Synaptic response of IGRT, based on the EDL effect, and its application to image recognition. a**, Illustrations of synaptic responses in biological neuronal networks and our ion-gating reservoir operating in an electric double layer mechanism. **b**, An example of the handwritten digit '6' from the MNIST database [44]. **c**, Drain current responses of the IGRT to 16 different pulse streams. **d**, Image recognition accuracy achieved by IGR as a function of the number of trained images. The dotted line shows the accuracy of a typical, full-simulation, 3-layer neural network (NN). The size of the IGR and NN networks are given in parentheses. The recognition accuracies of other physical reservoirs, such as Memristors (Magnetic skyrmion memristor (MSM) [13], $WO_x$ [16], $SiO_x$-Ag [18]) and Ionic liquid (IL) [43] are shown for comparison.

**Solving a second-order nonlinear dynamic equation by IGR**

Reservoir computing is suitable for time series data analysis because it possesses features such as short-term memory, nonlinearity, and high dimensionality for input data. We took advantage of such suitability by using the IGR to solve a second-order nonlinear dynamical equation task[13,16], a schematic of which is shown in Fig. 2a. The target $y_t(k)$ is obtained from following equation,

$$y_t(k) = 0.4y_t(k-1) + 0.4y_t(k-1)y_t(k-2) + 0.6u^3(k) + 0.1 \quad (1)$$

where $k$ and $u(k)$=[0,5] are a discrete time and a random input that were applied to IGRT as $V_G$ pulse streams, respectively. The reservoir states $X_i(k)$ were obtained from the $I_D$ response, and the reservoir output $y(k)$ is the linear combination of $X_i(k)$ and read out weights $w_i$ trained by ridge regression as follows,

$$y(k) = \sum_{i=1}^{N} w_i X_i(k) + b \quad (2)$$

where $N$ and $b$ are the reservoir size and bias respectively. Details of the procedure are given in the Methods section.

To obtain high-dimensional reservoir states from 1-dimensional input, IGRT with an 8-channel (drain)-1-gate-1-source structure and 8 different channel lengths (20 μm to 1000 μm) were fabricated as shown in Fig. 1a, and eight $I_D$ responses from common gate inputs were obtained as shown in Fig. 2b. The intensity of the spikes observed at the edges of $V_G$ pulses differ depending on the channel length. The spikes are due to a gate current induced by the ion current, which depends on the differential of charge in the EDL and are more significant with small $I_D$, compared to the gate current. Therefore, short channels with low resistance (≤100 μm) do not exhibit spikes, while long channels with high resistance (≥200 μm) exhibit large spikes. For further higher dimensionality of reservoir states, multiple reservoir states were obtained as virtual nodes, as shown in Fig. 2c [35]. The former nodes 1-5 and the latter nodes 6-10 correspond to $I_D$ responses measured during the application of write pulses and during the pulse intervals ($V_G$ =0 V), respectively. The former nodes utilize the fast relaxation process of channels from a low-resistivity state (LRS) to a high-resistivity state (HRS), which is dominated by $Li^+$ ion accumulation at the electrolyte/channel interface, while the latter nodes utilize a relaxation process from a HRS to a LRS of the channel, which is a relatively slow relaxation process because the $Li^+$ ion motion in the electrolyte is affected by channel resistance with HRS. Due to the ion-electron coupled dynamics, in which the ions of the electrolyte and the electrons of the channel interact, the IGRT exhibits asymmetric relaxation behavior. In addition, node 1 and node 6 are characteristic virtual nodes located at the peak of the spike-like drain current. Thus, by utilizing the virtual nodes, features such as asymmetric relaxation and spike behavior, which are unique features of the EDL, could be effectively extracted. These unique IGR features, induced by ion-electron coupled dynamics, will be discussed in Fig. 4. Fig. 2d shows the reservoir state obtained at each virtual node (1000 μm channel) for a random wave $u(k)$ input. It can be seen that the IGR has good diversity, with each virtual node showing various behaviors as reflections of its own characteristics. The combination of physical and virtual nodes resulted in a reservoir size of 80.

Fig. 2e shows the target for the test data and the predicted output by IGR. The predicted output is in excellent agreement with the target, that is, equation (1) was successfully solved by the IGR. As shown in Fig. 2f, the predicted error was $1.62 \times 10^{-4}$ (training data) and $2.08 \times 10^{-4}$ (test data), respectively. Compared to other physical reservoirs [13,16], the prediction error was extremely low, indicating that the IGR performs well on time series data analysis tasks. Such good computational performance of IGR is due to its ability to effectively exploit the complex and diverse features inherent in the ion-electron coupled dynamics of IGR as reservoir states. An additional important factor was the stable reproduction of the good expressivity of the IGR. This means that the nonlinear mapping to higher dimensional spaces by the IGR performed on the training data was exactly the same as for the test data, without altering the IGR condition during operation. This indicates that the IGR satisfies the echo-state-property (ESP), which is one of the important properties required for reservoirs [3].

**NARMA2 task**

We performed predictions on time series data generated by a NARMA2-system [45], as shown in equation (3), as a more challenging time series data analysis task. This is known as a NARMA2 task, and is commonly used as a typical reservoir computing benchmark task [7,34,36,37].

$$y_t(k+1) = 0.4y_t(k) + 0.4y_t(k)y_t(k-1) + 0.6u^3(k) + 0.1 \quad (3)$$

where $u(k)$=[0,0.5] is a random input. To evaluate the computational performance of IGR in the NARMA2 task, we used the normalized mean squared error (NMSE) for an index of RC performance, an explanation of which is given in the Methods section.

Fig. 3a shows the relationship between the IGRT operating conditions and the NMSEs (test phase) of the NARMA2 task. Good prediction performance was observed in the operation region with an input pulse period of 20 ms or longer and a duty ratio of 75% or higher. In particular, the best prediction performance (NMSE=0.023 in the test phase) was achieved at a pulse period of 50 ms and a duty ratio of 75%. The target and the predicted output by IGR (test phase) under these conditions are shown in Fig. 3b. Both waveforms are in excellent agreement, evidencing that IGR successfully predicted the time series generated by the NARMA2 system (please refer to Supplementary Information for details). Fig. 3c shows the relationship between the NMSE of the NARMA2 task in the test phase, and the volume of the physical reservoirs reported so far [7,34,36,37]. Although there are not many reports of physical reservoirs that experimentally demonstrate the NARMA task, IGR showed the best results in the prediction performance despite its extremely small volume compared to other physical reservoirs. That is, the IGR showed both extremely good computational performance on a single-device, and its suitability for integration.

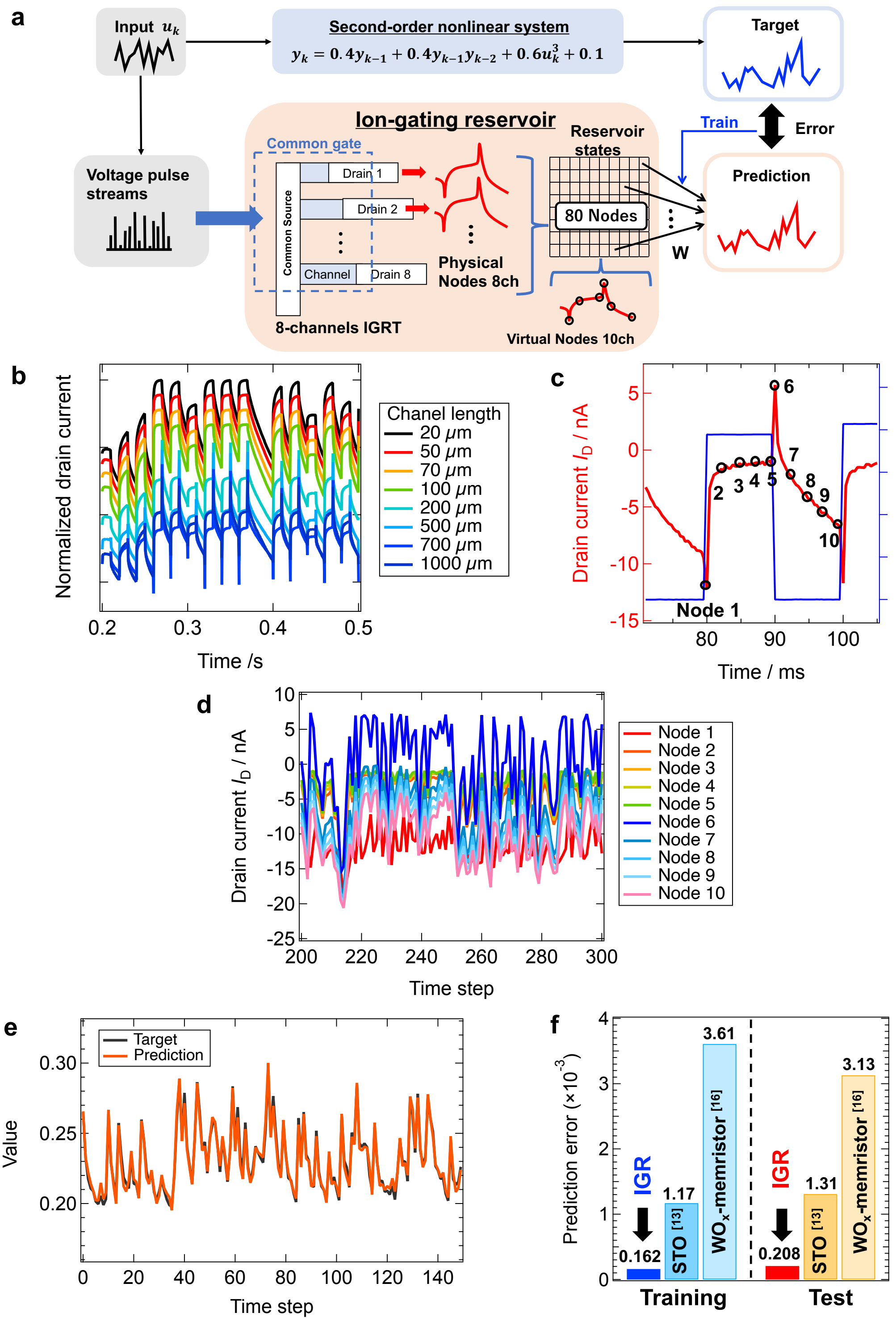

**Fig. 2 Solving a second-order nonlinear dynamic equation task. a**, Schematic of task calculated by IGR. **b**, Various drain current responses of the IGRT at different channel lengths. **c**, The method for obtaining virtual nodes and **d**, various reservoir states from 10 virtual nodes. **e**, Target and prediction waveforms of 2nd order nonlinear dynamic equation at the test phase. **f**, Prediction error compared to other physical reservoirs.

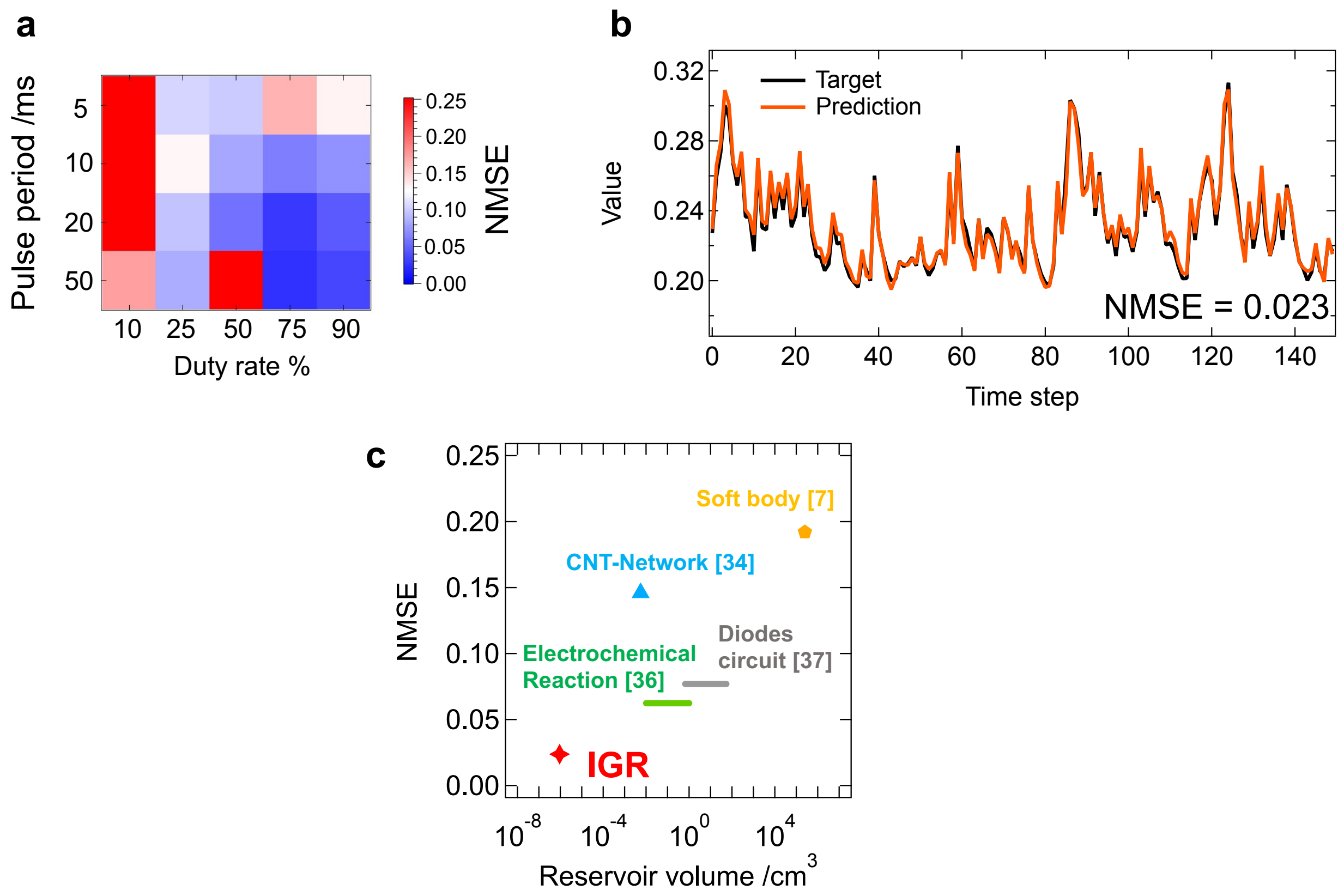

**Fig. 3 NARMA2 task. a**, The relationship between IGRT operating conditions and NMSEs in the test phase of the NARMA2 task. **b**, Target and prediction waveforms of the NARMA2 task. **c**, NMSEs of the NARMA2 task, and reservoir volumes of various physical reservoirs which experimentally demonstrated the NARMA2 task. The reservoir volume of IGR was calculated as the product of the total channel area and the thickness of the electrolyte.

## Simulation of ion-electron coupled dynamics in IGR

The ion and electron dynamics in our IGR were simulated by using COMSOL multiphysics simulation software (COMSOL, Inc.) in order to clarify the underlying mechanism in the unique *I*-*V* characteristics of our device. As shown in Fig. 4a, the EDLT model, which is comprised of a $Li^+$ electrolyte, a channel, and electrodes, was constructed by assuming the physical properties of LSZO, EDL, and the device structure. Please refer to the Methods section for details. Fig. 4b shows the $I_D$ response of the device model under four sequential gate pulse applied conditions. As seen in the rise and fall behavior of $I_D$, the model reproduces asymmetric $I_D$ responses with spikes that are signatures of our device, supporting our understanding that the simulation reproduces the actual electrochemical transport phenomena in the device. To grasp the ion and electron (hole) dynamics in the model in the operation, snap shots of the ion and hole density distribution are captured at specific points. In the snap shot at the initial state, shown in Fig. 4a, significant in-plane carrier distribution is found, in which densities of positively charged holes and negatively charged $Li^+$ vacancies are higher near the source electrode than near the drain electrode. This corresponds to formation of EDL, which is differently charged by voltage distribution due to application of $V_D$ (=-500 mV) between the source and drain electrodes. It is noted that the out-of-plane distribution of excess $Li^+$ (and $Li^+$ vacancies) accumulates within 0.3 nm from the interface. The extremely thin nature of the EDL is consistent with the in situ HAXPES observation [33]. Besides, comparison between the four conditions shown in Fig. 4c evidences that repetition of input makes a variety of charge density patterns in the channel. For example, at t=20 ms, not only the low hole density region proceeds from the drain side to the source side; an island-like pattern also appears within 0.3 nm from the channel/electrolyte interface. This is because such proceeding of the low hole density region occurs not only from the drain side but also from the source side, resulting in a variety of transient charge density patterns. Please refer to Supporting Information for a video of the charge density variation.

Basically, such behavior can be understood in the framework of transmission line model, in which electrical resistance is dependent on the location due to the different length of the current path [46]. However, in the present case, local hole resistance in the channel is strongly dependent on the charging history of EDL. This gives the $I_D$ response of the IGR asymmetric relaxation. Furthermore, spikes add another feature to the response. High performance of the IGR is discussed below as two contributions to the total drain current: $I_D$ to the source and $I_D$ to the gate.

Fig. 4d is an illustration of the $I_D$ path in the IGR. A partial $I_D$ (from the drain to source) corresponds to state variable $X(t)$ defined with an integral of local resistance $R(x,t)$ at each channel position $x$ $(0\leq x \leq L)$, $X(t) = \frac{V_D}{\int_{x=0}^{L} R(x,t)dx}$ , in which $L$, $x$, and $V_D$ are channel length, position in the channel, and drain voltage(constant), respectively. By introducing a local voltage applied to EDL $V_{EDL}(x,t)$ and EDL capacitance (constant) $C$ , $X(t)$ can be further transformed to

$$X(t) = \int_{x=0}^{L} V_D\, q\mu C V_{EDL}(x,t)dx \qquad (4),$$

in which $q$ and $\mu$ are charge and hole mobility, respectively. On the other hand, the rest of $I_D$ with a spike appearance (from the drain to the gate) corresponds to state variable $Y(t)$, defined with an integral of EDL charging current $I_{EDL}(x,t)$ at each channel position $x$, $(0\leq x \leq L)$,

$$Y(t) = \int_{x=0}^{L} I_{EDL}(x,t)dx = \int_{x=0}^{L} C\frac{dV_{EDL}(x,t)}{dt}dx \qquad (5),$$

in which $V_{EDL}(x,t)$ and $C$ are a local voltage applied to EDL and EDL capacitance (constant), respectively. As seen from equations (4) and (5), while both $X(t)$ and $Y(t)$ include $V_{EDL}(x,t)$, it is only expressed in a derivative form. Therefore, although both $X(t)$ and $Y(t)$ are related to $V_{EDL}(x,t)$, they function as two different types of reservoir. Since the drain current observed is the sum of $I_D$ to source and to gate, the $I_D$ response is a mixed reservoir of $X(t)$ and $Y(t)$, as shown in Fig. 4e. Recently, such mixed reservoirs have been theoretically predicted to show high performance by overcoming a trade-off relationship between short term memory and nonlinearity due to the coexistence, or mixture of linear dynamics and nonlinear dynamics in a reservoir [47]. The mixed reservoir property can be a reasonable explanation for the high performance discussed in time series data analysis tasks shown in Fig. 2 and Fig. 3.

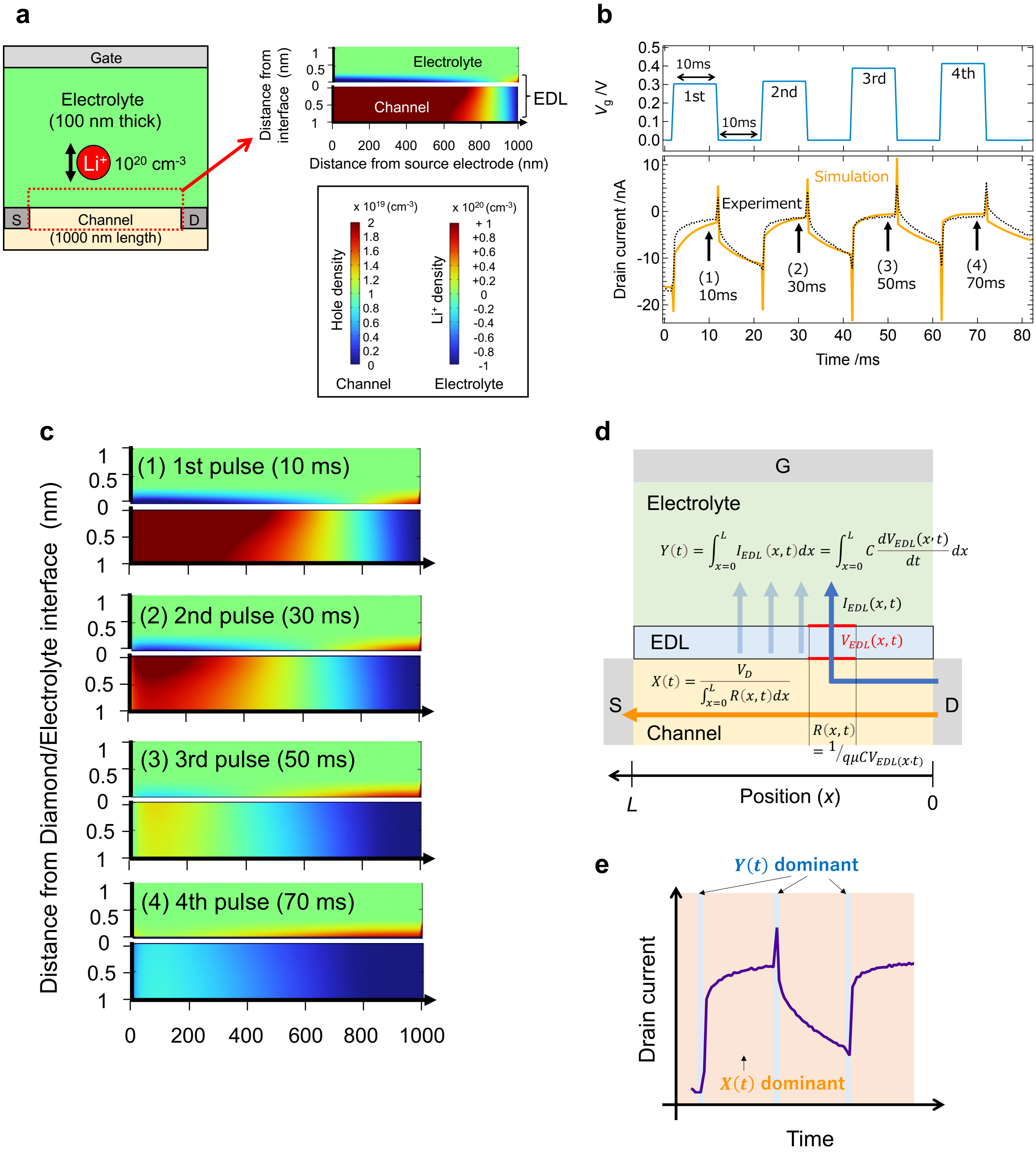

**Fig. 4 Simulation of ion-electron coupled dynamics. a**, The simulated EDLT, modelled by COMSOL Multiphysics, and the ion and hole distribution at the electrolyte/channel interface at the initial state. **b**, The drain current response of the simulated model under sequential gate pulse applied condition. The dotted line shows the experimental result. **c**, Snapshots of the ion and hole distribution, which are captured at each of the 4 pulses shown in **c**. **d**, Schematic illustration of the drain current path in the IGR, consisting of two drain currents; one corresponding to state variable *X*(*t*) and the other corresponding to state variable *Y*(*t*). **e**, The drain current response as a mixed reservoir of *X*(*t*) and *Y*(*t*).

**Lyapunov analysis**

To evaluate the high performance of the IGR in terms of nonlinear dynamics, we calculated the Lyapunov exponent, which quantifies the trajectory stability of the dynamical system by the Jacobi matrix method for direct analysis of time series data based on unknown dynamical systems [39,48]. Fig. 5a shows the nonlinear $I_D$ response used in the chaos time-series analysis for the 20 μm and 700 μm channels when a triangular wave is input to the IGRT. These channels exhibit completely different nonlinear responses, including the presence of spikes. To analyze the nonlinearity of IGR in detail, we generated 40 reservoir states $X$ by obtaining 5 virtual nodes, corresponding to Nodes 1 to 5 in Fig. 2c, for $I_D$ obtained from 8 channels.

Fig. 5b shows the return map ($X(k)$ vs $X(k+1)$) obtained from the reservoir states of nodes 1 and 5 with $L$=20 μm and $L$=20 μm ($X_{20\mu m,Node1}$, $X_{20\mu m,Node5}$, $X_{700\mu m,Node1}$, $X_{700\mu m,Node5}$). The return maps are completely different for each virtual node and for each channel length (physical node), which indicates that IGR achieves good diversity as a result of higher dimensioning by introducing virtual node and channel length. The return map at $L$=20 μm, shown in the upper panel of Fig. 5b, has a narrow trajectory width, indicating an almost completely periodic response to the triangular wave input. On the other hand, the return map at $L$=700 μm, shown in the lower panel of Fig. 5b, has a wide trajectory, indicating that the reservoir state has a relatively unstable response that varies slightly from period to period. Similar unstable characteristics have been reported for memristors [49] and nanowire networks [22] in chaos and edge-of-chaos states. We calculated Lyapunov exponents $\lambda$, an index of order-chaotic dynamics, of the IGR using the Jacobi matrix method [39,48]. Figure 5(d) shows the attractor in phase space created by selecting the axes in the $X_{20\mu m,Node1}$, $X_{700\mu m,Node1}$ and $X_{700\mu m,Node5}$ direction as one of the cross sections of the 41-dimensional phase space. The calculated Lyapunov spectrum is also shown in Fig. 6d. The Lyapunov exponents show values ranging from a minimum of -2.7 to a maximum of $8.3 \times 10^{-3}$: the maximum Lyapunov exponents $\lambda_{Max}$ of the IGR is $8.3 \times 10^{-3}$. Dynamical systems with maximum Lyapunov exponents $\lambda_{Max}$ near zero are called ‘edges of chaos’, and it has been reported that high computational performance is achieved at such edges of chaos in computing for physical reservoirs [14,22,24], full simulation reservoirs [38-40] and RNN [50], because of their robustness in the processing of information [25]. The high computational performance of IGR can also be attributed to edge-of-chaos, which was achieved by nonlinearity and high dimensionality. In other words, the high expressivity is realized by the asymmetric relaxation and spiking of the drain current, which originates from the ion-electron coupled dynamics at the electrolyte/semiconductor interface.

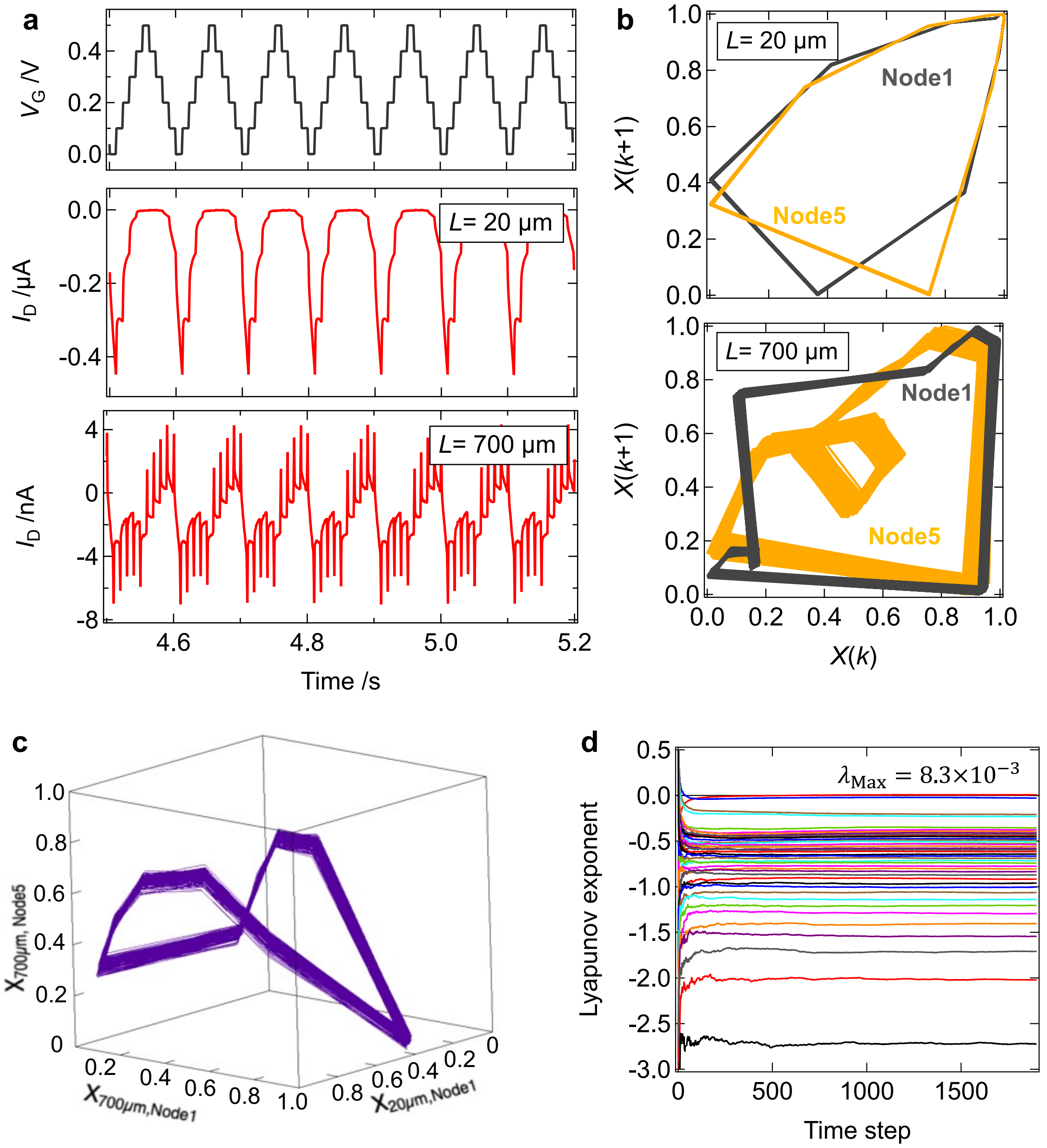

**Fig. 5 Lyapunov analysis. a**, Nonlinear drain current response of the IGR. Triangular wave input (upper panel) and drain current response of IGR obtained from 20 μm length channel (middle panel) and 700 μm length channel (lower panel). **b**, The return maps of the reservoir correspond to a 20 μm length channel (middle panel) and a 700 μm length channel (lower panel) with Node 1 and 5. **c**, The 3D cross section of the 41D reservoir state spaces of the IGR. **d**, Lyapunov spectrum of the IGR, calculated by the Jacobi matrix method.

## Conclusion

In order to achieve high performance reservoir computing (RC), an ion-gating reservoir (IGR) was developed, based on ion-electron coupled dynamics in the vicinity of a lithium ion solid electrolyte/diamond interface. In the study, various synaptic responses, with asymmetric relaxation and spikes, are effective in achieving excellent expression power for mapping time series data to higher dimensional feature space. Good RC performance of the IGR was demonstrated in handwritten digit recognition, nonlinear transformations, and NARMA tasks. Multiphysics simulation revealed that during operation, transient charge density patterns form and change from moment to moment in an extremely thin EDL region. Asymmetric relaxation and spikes in the drain current response enables high expression power by realizing a mixed reservoir comprising different nonlinear dynamics. Lyapunov analysis was performed to inspect the dynamical features of the IGR, which analysis revealed that the maximum Lyapunov exponents of the carrier dynamics is $8.3 \times 10^{-3}$, supporting the understanding that the IGR operates in 'edge of chaos' states under certain conditions. Furthermore, the concept of an ion-gating reservoir can be extended to various information carriers (e.g., electrons, ions, light, and spin) as long as their dynamics or transport interact with each other. While the present EDL system, with its ion-gating transistor structure, is a typical case, various physical or chemical systems can be utilized for achieving IGR with diverse information carriers. Various materials and interfaces present exciting frontiers for exploring high performance, versatile, and integrated physical RC based on the coupled dynamics inherent in IGR.

# Methods

## Device fabrication

Hydrogen-terminated diamond was deposited on a single crystal diamond substrate (100) (EDP) by the MPCVD method. During deposition, 500 sccm and 0.5 sccm of $H_2$ and $CH_4$, respectively, were introduced and the hydrogen-terminated diamond was grown at 950 W RF power. The IGRT were fabricated with eight different channel lengths (20 μm, 50 μm, 70 μm, 100 μm, 200 μm, 500 μm, 700 μm, 1000 μm,), all with a channel width of 100 μm. Pd/Pt electrodes (10 nm and 35 nm, respectively) were deposited by electron beam evaporation with maskless lithography after oxygen termination of the diamond surfaces, other than channels, by oxygen plasma asher. A 3.5 μm LSZO thin film, used as an electrolyte, was deposited by PLD with an ArF excimer laser. 100 nm of $LiCoO_2$ was deposited by PLD to form the gate electrode, and a 50 nm Pt thin film was deposited by electron beam deposition.

## Measurement method

IGRT measurements were carried out at room temperature in a vacuum chamber evacuated by a turbo molecular pump. Probers were used to connect the IGRT in the chamber, and electrical measurements were performed using the source measure unit (SMU) and pulse measure unit (PMU) of a semiconductor parameter analyzer (4200A-SCS, Keithley).

## Image recognition

A handwritten digit from the MNIST dataset [44] was used for the task of image recognition by IGR, with 60,000 images as training data and 10,000 as test data. Each pixel of a 28x28 pixel handwritten digit was converted to a binary state of "0" or "1" and input to the IGRT as a time-series data signal every 4 pixels. "0" and "1" were set to $V_G$=-0.5 V and $V_G$ =0.3 V, respectively, and the pulse widths were 10 ms for each. The pulse interval was set to 12.5 ms, and $V_G$ =-0.5 V was also applied during this time. While measuring the drain current by applying a constant drain voltage ($V_D$ =-0.5 V) to the IGRT, 16 different time-series data ("0000"~"1111") were input to the IGRT, and a drain current 12.5 ms after the fourth pulse input was obtained as the reservoir state. The 16 drain currents obtained by the measurements were normalized so that the drain current corresponding to "1111", which has the largest value, became 1 after adding 6 μA as an offset to the drain current. Then, the 784-pixel digit data was replaced with the reservoir state every 4 pixels, and the digit data converted to 196 values of reservoir state was trained and classified in the readout network. With the reservoir state matrix as $\boldsymbol{X}$ and the weight matrix of the readout network as $\boldsymbol{W}$, the readout function is defined as follows

$$h(\boldsymbol{X}) = g(\boldsymbol{W} \cdot \boldsymbol{X}) \quad (6),$$

$$g(\boldsymbol{z}) = \frac{1}{1 + e^{-\boldsymbol{z}}} \quad (7).$$

The squared error is defined as

$$E = \frac{1}{2}\sum_{i=1}^{10}\left(y_i - h(x_i)\right)^2 \quad (8).$$

The weights $W$ were updated to minimize

$$\Delta W = -\alpha \frac{\partial E}{\partial W} \quad (9)$$

The learning rate α was set to 0.1, and the training was performed 20 times.

**Solving second-order-nonlinear dynamic equation tasks**

A random input $u(k)$ was converted to voltage pulse streams, with a pulse width of 10 ms and an interval of 10 ms, and applied to the gate terminal of the IGRT. The intensity of the pulses $V_G(k)$ was equal to $u(k)$, over a range of from 0 to 0.5 V, and $V_G$ =0 V was applied during pulse intervals. The drain current responses of the IGRT with 8 channels were measured under constant $V_D$=-0.5 V, and 10 virtual nodes were obtained from each of them. Thus, 80 reservoir states were obtained from 1D input $u(k)$ by IGR. These reservoir states were normalized from 0 to 1 for calculation, as shown in equation (2).

**Ridge regression for solving second-order nonlinear dynamic equation tasks and NARMA2 tasks**

In the time series data analysis tasks shown in Fig. 2 and Fig. 3, the readout network of IGR was trained by ridge regression. Here we explain the algorithm for ridge regression. The reservoir output $y(k)$ shown in equation (2) is transformed to

$$y(k) = \boldsymbol{W} \cdot \boldsymbol{X}(k) \quad (10),$$

where $\boldsymbol{W} = (w_0, w_1, \ldots, w_N)$ and $\boldsymbol{X}(k) = \left(X_0(k), X_1(k), \ldots, X_N(k)\right)^{\mathrm{T}}$ are the weight vector and the reservoir state vector with a reservoir size of $N$, respectively. Note that $w_0 = b$ and $X_0(k) = 1$ to introduce the bias $b$ shown in equation (2). The cost function $J(\boldsymbol{W})$ in ridge regression is defined as follows

$$J(W) = \frac{1}{2}\sum_{k=1}^{T}\left(y_{\mathrm{t}}(k) - y(k)\right)^2 + \frac{\lambda}{2}\sum_{i=0}^{N} w_i^2 \quad (11),$$

where $T$, $\lambda$ and $y_{\mathrm{t}}(k)$ are the data length in the training phase, the ridge parameter and the target output generated by equation(1) or equation(3), respectively. We fixed $T = 450$ and $\lambda = 5 \times 10^{-4}$ for all the tasks demonstrated in Fig. 2 and Fig. 3. The weight matrix $\widehat{\boldsymbol{W}}$ that minimizes cost function $J(\boldsymbol{W})$ is given by following equation (12)

$$\widehat{\boldsymbol{W}} = \boldsymbol{Y}\boldsymbol{X}^{\mathrm{T}}(\boldsymbol{X}\boldsymbol{X}^{\mathrm{T}} + \lambda\boldsymbol{I})^{-1} \quad (12),$$

where $\boldsymbol{Y} = \left(y_t(1), y_t(2), \ldots, y_t(T)\right)$, $\boldsymbol{X} = \left(\boldsymbol{X}(1), \boldsymbol{X}(2), \ldots, \boldsymbol{X}(T)\right)$ and $\boldsymbol{I}\left(\subseteq \mathbb{R}^{(N+1)\times(N+1)}\right)$ are the target output vector, the reservoir state matrix and the identify matrix, respectively.

Then, after learning the readout weights, the computational performance was evaluated by 'Prediction error' for solving the second-order-nonlinear dynamic equation task and 'NMSE' for the NARMA2 task, as shown in following equation (13,14)

$$\text{Prediction error} = \frac{\sum_{k=1}^{T}\left(y_t(k) - y(k)\right)^2}{\sum_{k=1}^{T} y_t^2(k)} \quad (13),$$

$$\text{NMSE} = \frac{1}{T}\frac{\sum_{k=1}^{T}\left(y_t(k) - y(k)\right)^2}{\sigma^2(y_t(k))} \quad (14),$$

where $T$ is a data length in the training phase ($T = 450$) or test phase ($T = 150$).

**COMSOL Multiphysics**

A simulation of IGRT was performed with the semiconductor module of the CMSOL multiphysics software application, so as to analyze the dynamical behavior of ions and electrons in IGRT. The model of IGRT simulated by COMSOL consists of the following three parts: (1) $Li^+$ electrolyte with Li concentration $10^{22}$ $cm^{-3}$ and mobility $4 \times 10^{-13}$ $cm^2/Vs$ (100 nm thick). (2) EDL with constant capacitance of 4.0 μF/$cm^2$. (3) diamond channel with hole concentration $10^{19}$ $cm^{-3}$ and mobility 150 $cm^2/Vs$ (1 nm thick and 1 μm channel length). These parts are modeled in two-dimensional space in the source-drain and channel-gate directions, and effects in the depth direction (i.e., channel width direction) are considered only for current scaling.

**Lyapunov analysis**

Here, we introduce how the Lyapunov exponents of IGR were calculated by the Jacobi matrix method [39,48]. The Jacobi method matrix estimates the Jacobi matrix from points on the attractor in the m-dimensional phase space of the time series data. Let us consider an m-dimensional sphere ($\epsilon$-sphere) of infinitesimal radius $\epsilon$, centered at a point $\boldsymbol{v}(t)$ on the orbit of the attractor at time $t$. With $\boldsymbol{v}(k_i)$ $(i = 1,2,\dots,M)$ as the other points on the attractor located inside the $\epsilon$ -sphere, the displacement vector $\boldsymbol{\mu}_i$ of $\boldsymbol{v}(k_i)$ as seen from $\boldsymbol{v}(t)$ is obtained as follows;

$$\boldsymbol{\mu}_i = \boldsymbol{v}(k_i) - \boldsymbol{v}(t) \quad (15).$$

Also, the displacement vector $\boldsymbol{z}_i$ after time $s$ is obtained as follows, and a small radius of $\epsilon$ -sphere and $s$ allows a linear approximation shown on the right of the equation as $\hat{\boldsymbol{J}}(t)$ which is the Jacobi matrix to estimate,

$$\boldsymbol{z}_i = \boldsymbol{v}(k_i + s) - \boldsymbol{v}(t + s) \approx \hat{\boldsymbol{J}}(t)\boldsymbol{\mu}_i \quad (16).$$

Then, from equation (16), the Jacobi matrix can be estimated as follows,

$$\hat{\boldsymbol{J}}(t) = \boldsymbol{z}_i\boldsymbol{\mu}_i^{\mathrm{T}}\left(\boldsymbol{\mu}_i\ \boldsymbol{\mu}_i^{\mathrm{T}}\right)^{-1} \quad (17).$$

Considering the QR decomposition of the Jacobi matrix, the Jacobi matrix at time t can be expressed as shown in equation (18), and the Lyapunov exponent $\lambda_i$ for $i = 1,2,\dots,m$ is calculated as shown in equation (19),

$$\hat{\boldsymbol{J}}(t)\boldsymbol{Q}_t = \boldsymbol{Q}_{t+1}\boldsymbol{R}_{t+1} \quad (18),$$

$$\lambda_i = \lim_{T\to\infty}\frac{1}{2T}\sum_{k=1}^{2T}\log\left|\boldsymbol{R}_k^{ii}\right| \quad (19),$$

where $\boldsymbol{R}_k^{ii}$ is the $i$-th diagonal element of the upper triangular matrix $\boldsymbol{R}_k$.

**Acknowledgement:** This work was in part supported by Japan Society for the Promotion of Science (JSPS) KAKENHI Grant Number JP22H04625 (Grant-in-Aid for Scientific Research on Innovative Areas "Interface Ionics"), and JP21J21982 (Grant-in-Aid for JSPS Fellows). A part of this work was supported by the Yazaki Memorial Foundation for Science and Technology and Kurata Grants from The Hitachi Global Foundation.

**Author contributions:** T.T., D.N., W.N., and K.T. conceived the idea for the study. D.N., T.T., and W.N. designed the experiments. D.N. and T.T. wrote the paper. D.N. and W.N. carried out the experiments. D.N., W.N., and M.T. prepared the samples. D.N., T.T., and W.N. analyzed the data. D.N. carried out the multi-physics calculations. All authors discussed the results and commented on the manuscript. K.T. directed the projects.

**Competing interests:** The authors declare that they have no competing interests.

**Date and materials availability:** All of the data needed to evaluate the conclusions in the paper are present in the paper and/or the Supplementary Materials. Additional data related to this paper may be requested from the authors.

## Supplementary information

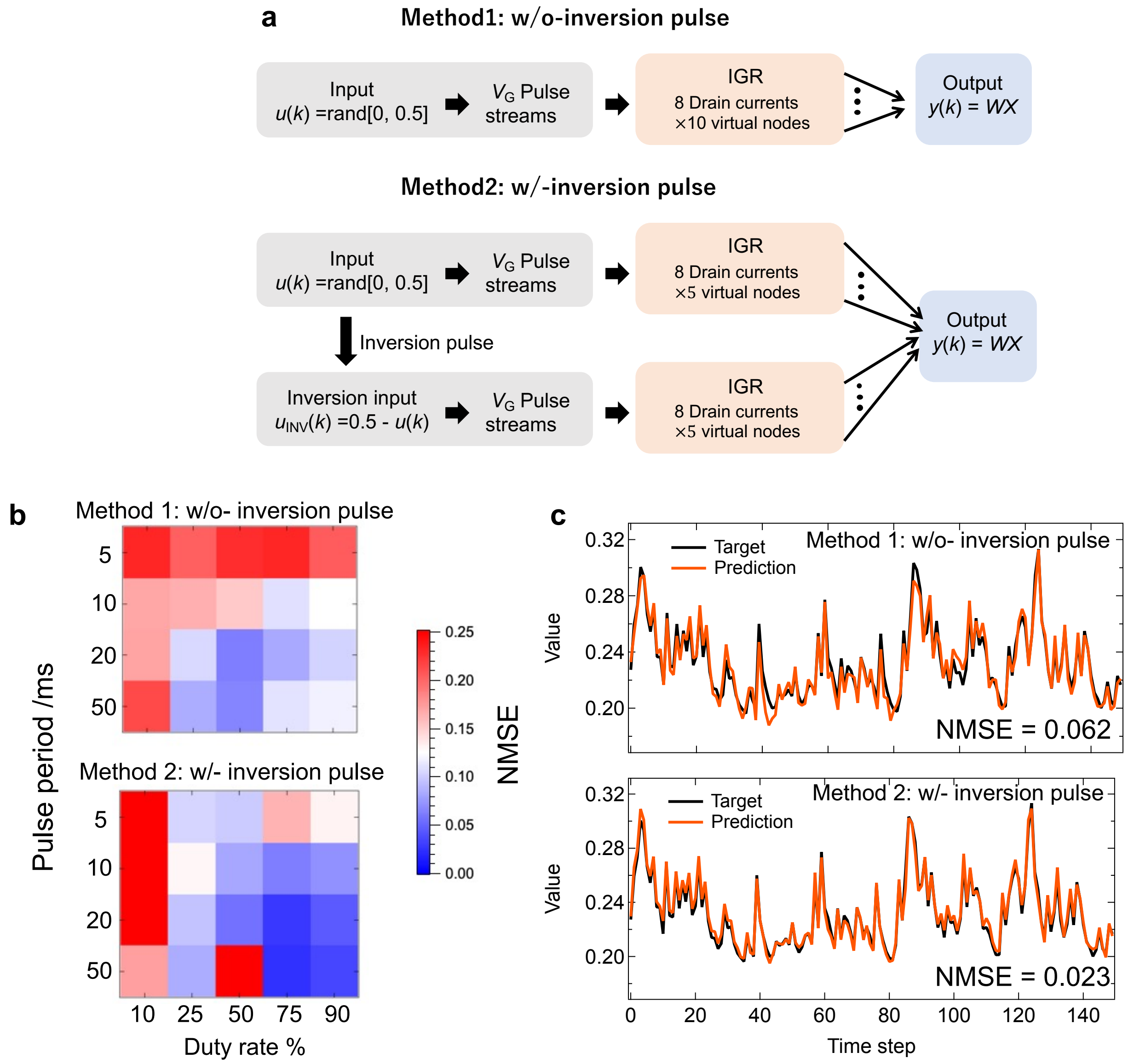

**Supplementary Fig. 1 NARMA2 task demonstrated with 2 different input methods by IGR. a**, Schematic illustration of Method 1 (w/o-inversion pulse) and Method 2 (w/-inversion pulse). **b**, Relationship between IGRT operating conditions and NMSEs in the test phase of the NARMA2 task in Method 1 (upper panel) and Method 2 (lower panel). **c**, Target and prediction waveforms of the NARMA2 task in Method 1 (upper panel) and Method 2 (lower panel).

### NARMA2 task demonstrated with 2 different input methods by IGR

We performed the NARMA2 task with two input methods, as shown in Supplementary Fig. 1a. Method 1 is the same procedure as in the second-order nonlinear dynamic equation task, where u(k) is converted to a pulse voltage signal $V_G(k)$ and input to the device (w/o-inversion pulse). Increasing physical nodes is effective for high performance RC. In addition to the normal reservoir states obtained in Method 1, Method 2 uses additional reservoir states, which are obtained by inversion pulses. The procedure is as follows; $u_{\mathrm{inv}}(k)$ with the intensity of $u(k)$ inverted as shown in equation (1) was converted to a pulse voltage signal $V_{\mathrm{inv}}(k)$ and input to the IGRT, and an additional reservoir state was obtained by the virtual node method from an additional 8 drain current responses (w/-inversion pulse). To compare these methods without any contribution from differences in network size, the total number of reservoir states for Method 1 and 2 were kept to 80 nodes. Fig. 3 shows the results from Method 2.

$$u_{\mathrm{inv}}(k) = 0.5 - u(k) \tag{1}$$

The upper panel of Supplementary Fig. 1b shows the relationship between the IGRT operating condition and the NMSEs (test phase) of the NARMA2 task in Method 1 (wo/-inversion pulse). Good prediction performance was observed in the operation region with an input pulse period of 20 ms or longer and a duty ratio of 25% or higher. In particular, the best prediction performance (NMSE=0.062 test phase) was achieved at a pulse period of 20 ms and a duty ratio of 50%. The target waveform and the predicted waveform of the IGR during the test phase under this condition are shown in the upper panel of Supplementary Fig. 1c. These two waveforms are in good agreement, indicating that IGR successfully predicted the time series generated by the NARMA2 system. The lower panel in Supplementary Fig. 1b shows the relationship between the IGRT operating conditions and the NMSEs (test phase) of the NARMA2 task in Method 2 (w/- inversion pulse). Compared to Method 1, the prediction performance improved overall, with good prediction performance in the operation region with an input pulse period of 20 ms or longer and a duty ratio of 75% or higher. In particular, the best prediction performance (NMSE=0.023 in test phase) was achieved at a pulse period of 50 ms and a duty ratio of 75%. The target waveform and the predicted waveform of IGR (test phase) under these conditions are shown in the lower panel of Supplementary Fig. 1c. Both waveforms are in excellent agreement.

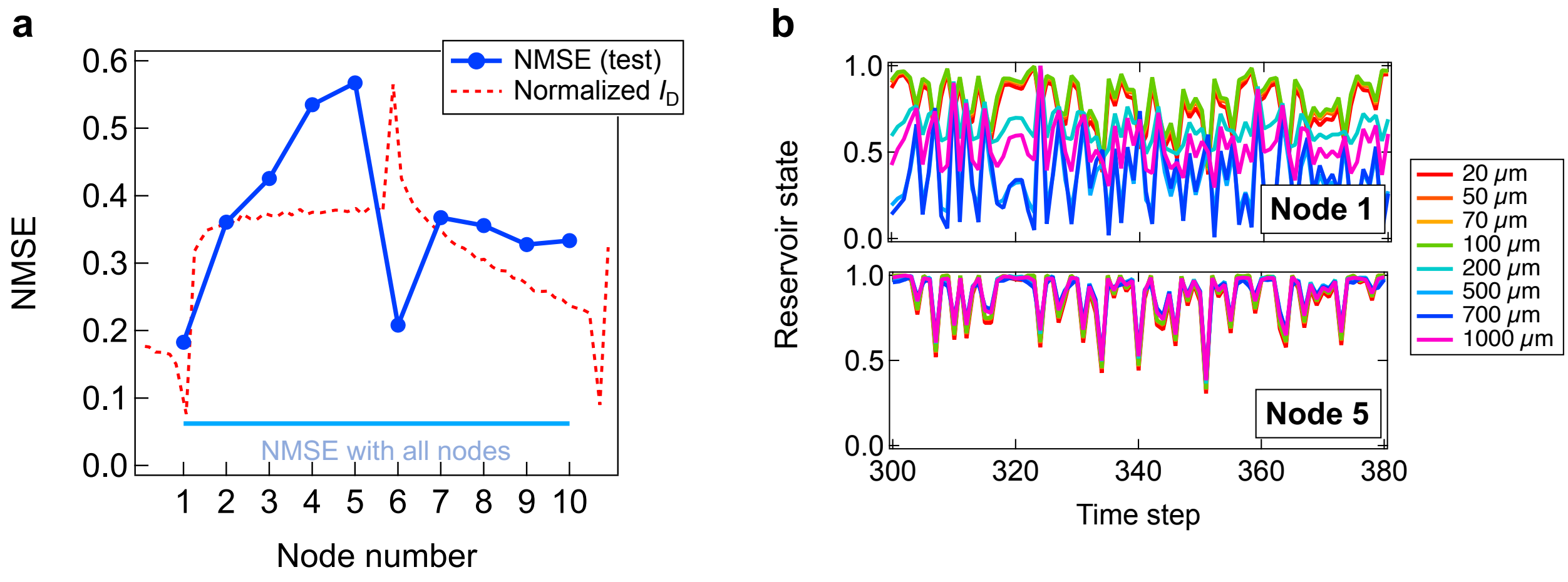

**Supplementary Fig. 2 Virtual node dependence in IGR computation performance a**, Virtual node dependence of NMSE in the NARMA2 task. The dotted lines indicate the normalized drain current, done so that the obtained virtual node location corresponds to the NMSE. **b**, Reservoir states obtained from virtual node 1(upper panel) and 5 (lower panel) correspond to Fig. 2c.

### Virtual node dependence in IGR computation performance

To investigate the origin of such high computational performance of the IGR, we evaluated the virtual node dependence of NMSE for the NARMA2 task shown in Supplementary Fig. 2a. Only physical nodes without inversion pulses were used here to predict the NARMA2 system (8 reservoir states). The blue line in the figure shows the NMSE with all nodes (80 reservoir states). The best prediction performance was obtained for virtual nodes 1 and 6, which correspond to the spike behavior of the drain current, as shown in Supplementary Fig. 2a. This indicates that such spikes are not just noise, but also contribute significantly to the computation. It has in fact been reported that the effect of noise in RC is to reduce computational performance [1]. As shown in the top panel of Supplementary Fig. 2b, the reservoir state of virtual node 1, which had the best computational performance, is rich in diversity and nonlinearly transforms the input signal to higher dimensions, while the reservoir state of virtual node 5, which had the worst performance, showed relatively low diversity. The highly efficient computation of IGR is achieved by said virtual and physical nodes, which effectively extract the features of the input signal through the EDL effect.

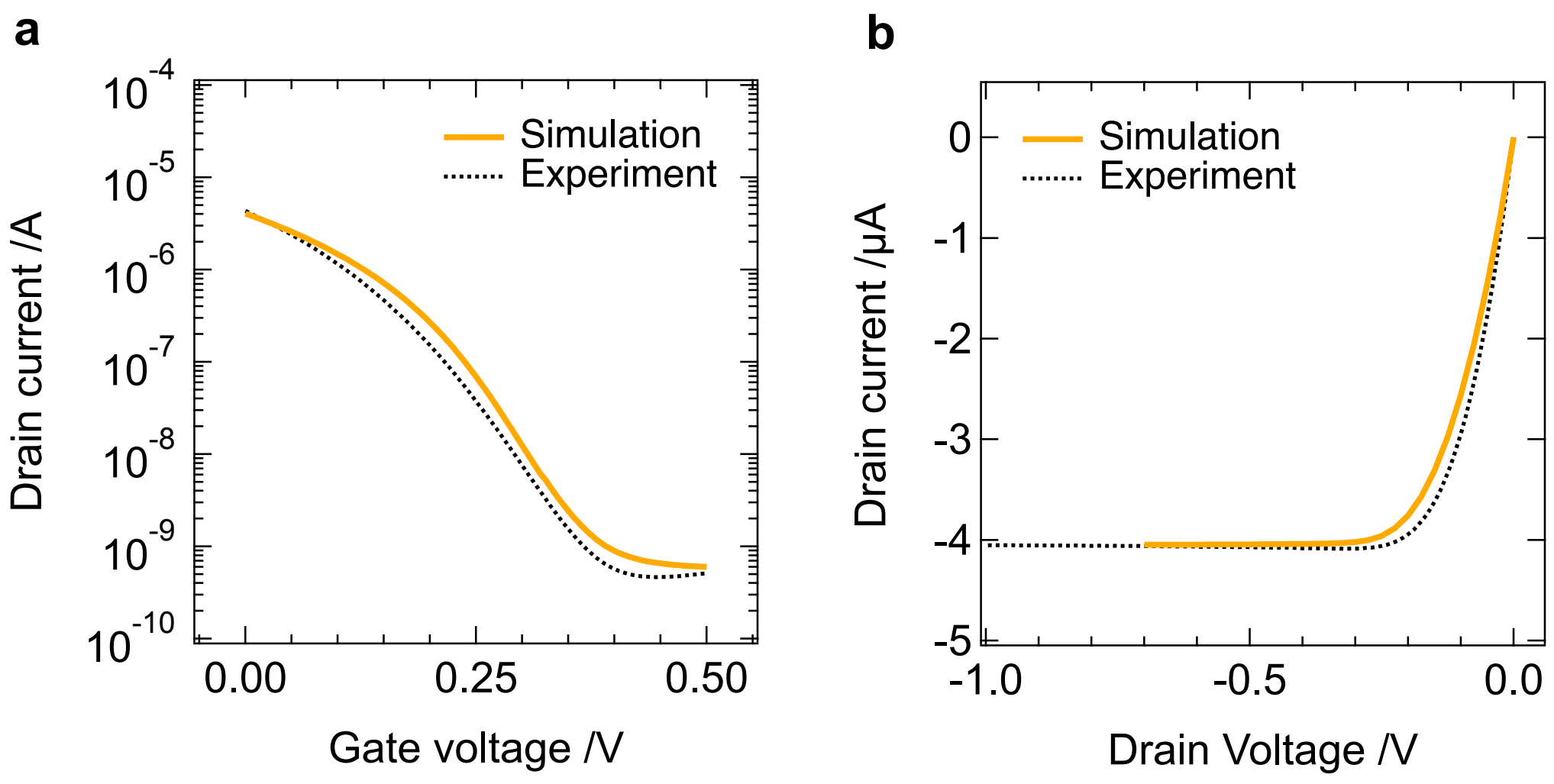

**Supplementary Fig. 3 a**, $I_D$-$V_G$ and **b**, $I_D$-$V_D$ characteristics of the simulated EDLT model. The dotted line shows the experimental result by IGRT.

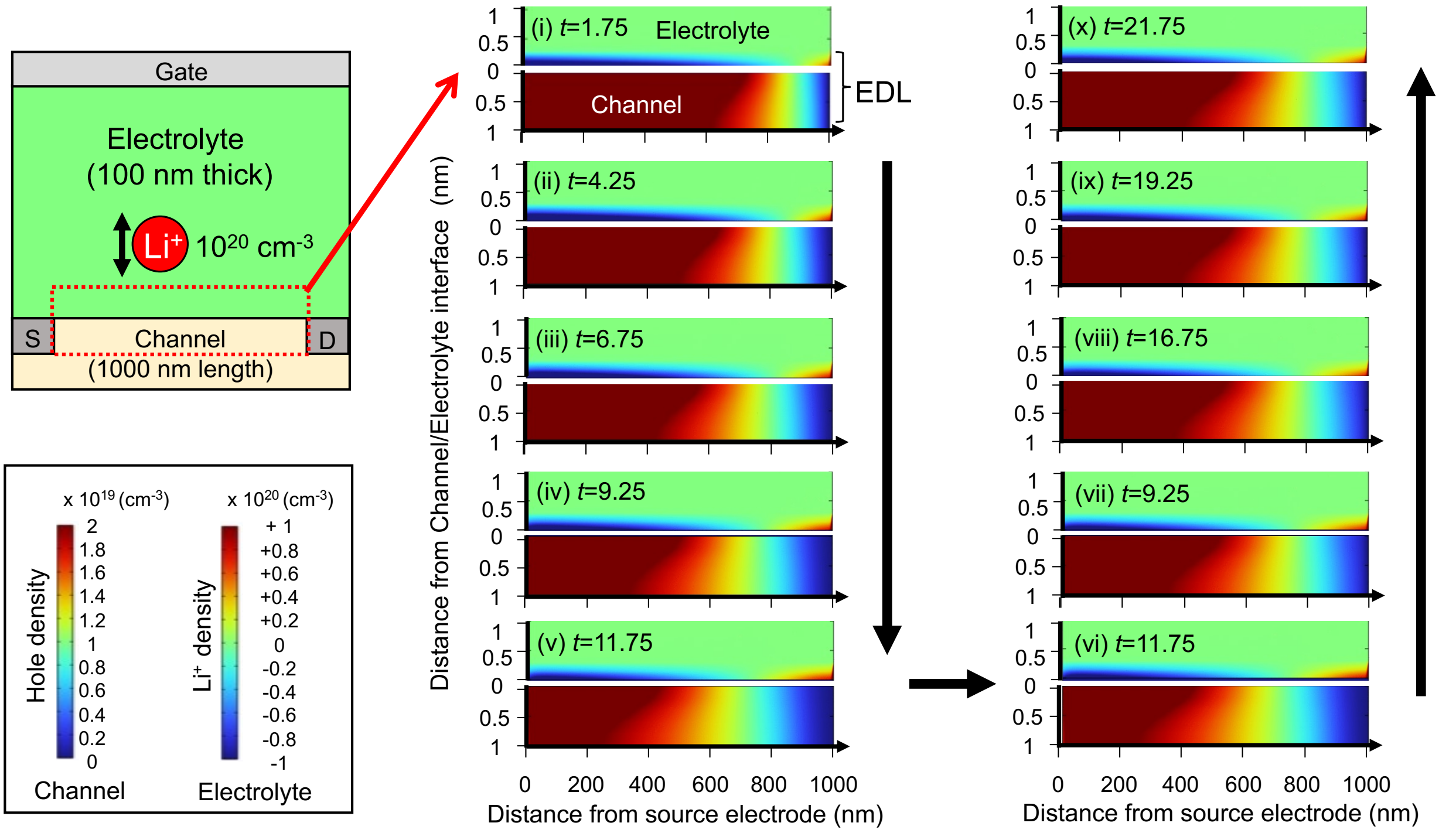

**Supplementary Fig. 4** The ion and hole distribution at the electrolyte/channel interface during the 1st pulse operation in Fig. 4b.

**Simulation of ion-electron coupled dynamics in IGR**

The ion and electron dynamics in our IGR were simulated using the COMSOL multiphysics software application in order to clarify the underlying mechanism in the unique *I-V* characteristics of our device. The EDLT model, which is comprised of $Li^+$ electrolyte, channel and electrodes, was constructed by assuming the physical properties LSZO($Li^+$ concentration: $10^{22}$*$cm^{-3}$, mobility: $4 \times 10^{-13}$ $cm^2/Vs$), EDL(constant capacitance: 4.0 μF/$cm^2$), and the device structure, some of which were modified to reduce computational load. Please refer to the Method section for details of the calculation. After tuning the device parameters, the $I_D$-$V_G$ and $I_D$-$V_D$ characteristics of the simulated model shown in Supplementary Figs. 3a,b agreed well with the experimental result shown for reference.

In the initial state [$t$=1.75 ms indicated as (i) in Supplementary Fig. 4] of both channel (corresponds to diamond channel of the EDLT) and electrolyte (corresponds to LSZO of the EDLT), significant in-plane carrier distribution was found, in which the densities of positively charged holes and negatively charged $Li^+$ vacancies are higher near the source electrode than near the drain electrode. This corresponds to the formation of EDL, which is differently charged by voltage distribution due to application of $V_D$(=-500 mV) between the source and drain electrodes. It is noted that any out-of-plane distribution of excess $Li^+$(and $Li^+$ vacancy) accumulates within 0.3 nm from the interface. The extremely thin nature of the EDL is consistent with the result of in situ HAXPES observation of LSZO/Au interface [2]. During application of the first $V_G$ pulse [from $t$=2 to 12 ms, (ii) to (v) in Supplementary Fig. 4], $Li^+$ moves towards the channel/electrolyte interface, the $Li^+$ accumulated region in the electrolyte and the hole depletion region in the channel gradually proceed from the drain electrode to the source electrode. When the opposite occurs in $Li^+$ density and hole density, $V_G$-driven EDL charging occurs in the vicinity of the channel/electrolyte interface through electrostatic interaction between negatively charged $Li^+$ vacancies and positively charged holes. After removal of the 1st pulse [from $t$=12 to 22 ms, (vi) to (x) in Supplementary Fig. 4], the reverse process takes place in both the electrolyte and the channel. The comparison between (i) the initial state ($t$=1.75 ms) and a long time period after the removal of the 1st pulse (x) ($t$=21.75 ms) in Supplementary Fig. 4 evidences that the carrier distributions are in two different states, meaning that reverse processes (12 ms$<t<$22 ms) have longer relaxation times than forward process (2 ms$<t<$12 ms), which provides good short term memory to store the input history. This is consistent with the drain current response shown in Fig. 4b.